\documentclass[%
 amsmath,amssymb,
 aip,
 jcp,
 reprint,
 twocolumn,
]{revtex4-2}

\usepackage{graphicx}
\usepackage{dcolumn}
\usepackage{bm,color}
\usepackage{amsmath, amssymb, amsthm}
\usepackage{hyperref}

\begin{document}

\preprint{APS/123-QED}

\title{Large deviations of ionic currents in dilute electrolytes}

\author{Jafar Farhadi}
\affiliation{Department of Chemical and Biomolecular Engineering, University of California, Berkeley, CA, USA \looseness=-1}
\author{David T. Limmer}
\email{dlimmer@berkeley.edu}
\affiliation{Department of Chemistry, University of California, Berkeley, CA, USA}
\affiliation{Kavli Energy NanoScience Institute, Berkeley, CA, USA}
\affiliation{\mbox{Materials Sciences Division, Lawrence Berkeley National Laboratory, Berkeley, CA, USA}}
\affiliation{\mbox{Chemical Sciences Division, Lawrence Berkeley National Laboratory, Berkeley, CA, USA}}

\date{\today}

\begin{abstract}
We evaluate the exponentially rare fluctuations of the ionic current for a dilute electrolyte by means of macroscopic fluctuation theory. We consider the fluctuating hydrodynamics of a fluid electrolyte described by a stochastic Poisson-Nernst-Planck equation. We derive the Euler-Lagrange equations that dictate the optimal concentration profiles of ions conditioned on exhibiting a given current, whose form determines the likelihood of that current in the long-time limit. For a symmetric electrolyte under small applied voltages, number density fluctuations are small, and ionic current fluctuations are Gaussian with a variance determined by the Nernst-Einstein conductivity. Under large applied potentials, where number densities vary, the ionic current distribution is generically non-Gaussian. Its structure is constrained thermodynamically by Gallavotti-Cohen symmetry and the thermodynamic uncertainty principle.
\end{abstract}

\maketitle

\section{\label{sec:main1}Introduction}
Transport processes on nanoscales are fundamentally different than those that occur macroscopically. On small scales, applied gradients can be large such that linear response approximations break down,\cite{limmer2021large} leading to driving-force-dependent transport coefficients\cite{onsager1957wien,toquer2025ionic,gao2019nonlinear} or modified constituent relationships.\cite{lepri2011density,ray2019heat,chen2008nanoscale} Further, when device feature sizes are small and driving forces are comparable to thermal energies, fluctuations about their mean behavior can be substantial such that it is important to understand the distribution of dynamical events, not just their typical values.\cite{camalet2003current,krems2009effect,helms2025stochastic,kim2023electrical,marbach2021intrinsic} Unlike configurational statistics within thermal equilibrium, there is no generic form for the distribution of dynamic quantities. However, in the limit of long observation times, large deviation theory offers a means of evaluating the distribution of time-integrated dynamical quantities.\cite{touchette2018introduction,limmer2024statistical} In this scaling limit, macroscopic fluctuation theory provides a route to the distribution function by solving an optimization problem.\cite{bertini2015macroscopic} Here we apply macroscopic fluctuation theory to a dilute electrolyte and evaluate the distribution of ionic currents resulting from an electrostatic potential drop.  These calculations offer a set of analytical and numerically exact results with which to begin to understand the far-from-equilibrium fluctuations of electrolytes.

Detailed observations on synthetic nanofluidic devices and naturally occurring ion transport channels are increasingly challenging our macroscopic understanding of fluid transport.\cite{kavokine2021fluids,gouaux2005principles} Often, the working fluids are simple aqueous electrolytes whose bulk properties are well known, and yet when confined to nanometer-scale channels and driven far from equilibrium, their emergent behaviors are difficult to anticipate.\cite{bocquet2020nanofluidics} Theory and computer simulations have been used to understand typical fluctuations around nonequilibrium steady-states generated by applied voltages, fields, and concentration gradients.\cite{rotenberg2013electrokinetics,daiguji2010ion,peters2016analysis} These studies have uncovered the role of boundary conditions in affecting observed conductivities and selectivities of channels,\cite{joly2006probing,joly2021osmotic,poggioli2021distinct,helms2023intrinsic,huang2008water} and provided some design principles for novel nonlinear devices like ionic diodes and memristors.\cite{lee2016tuning,gao2014high,jubin2018dramatic}  

A few computer simulations have been performed using advanced simulation tools\cite{ray2018exact,gao2017transport,singh2025variational} to probe exponentially rare current fluctuations in bulk electrolyte solutions.\cite{lesnicki2021molecular,lesnicki2020field} However, an understanding of the fluctuations about the typical behavior is largely lacking since a formalism to predict the form of such distributions has not been developed. This renders the conclusions of the few simulations done difficult to generalize to other systems or to boundary conditions that are not easily studied computationally. Here, we bridge this knowledge gap by establishing some results for a simple, dilute electrolyte where analytical progress can be made and numerically exact results obtained.

The theory of dynamical large deviations has established a formal means of evaluating and interpreting fluctuations of time-extensive quantities.\cite{jack2020ergodicity} Together with stochastic thermodynamics, universal symmetries and scaling relations for fluctuations of time-integrated currents have been established.\cite{seifert2012stochastic} For example, the relative likelihoods of observing currents in one direction or its opposite are given by fluctuation theorems.\cite{sevick2008fluctuation,jarzynski2011equalities} Generically, the time-scaled log-likelihoods of time-extensive currents are themselves time independent and obey Gallavotti-Cohen symmetry.\cite{gallavotti1995dynamical} While large deviation theory was initially restricted to the domain of abstract formal results or applications to idealized systems, increasingly it has been used to understand fluctuations of molecularly detailed systems with advanced computer simulation techniques and theoretical developments.\cite{limmer2021large,jack2020ergodicity,singh2025variational}

Macroscopic fluctuation theory is a theoretical tool for evaluating current fluctuations within nonequilibrium steady-states. By formulating a stochastic hydrodynamic equation of the relevant currents and corresponding density fields, macroscopic fluctuation theory casts the evaluation of the current distribution function into an optimization problem.\cite{bertini2002macroscopic} In the long-time limit, the likelihood of a specific current fluctuation is determined by the least unlikely, or optimal, density profile that generates that current.\cite{bodineau2004current} Optimal profiles for continuum limits of lattice models of heat and mass transport have been evaluated, but the application of this formalism to electrolyte solutions and their ionic transport is missing.\cite{hurtado2009test,agranov2023macroscopic,PhysRevE.87.032115,hurtado2025optimal} Applying macroscopic fluctuation theory to a strong electrolyte confined to a one-dimensional channel, we find that the mechanism of rare current fluctuations depends on the potential drop. Near equilibrium, when applied potentials are small, rare current fluctuations are generated by density fluctuations that are delocalized throughout the channel. For large applied potentials, rare current fluctuations are generated by the formation of boundary layers, gradients in the density exponentially close to the ends of the channel. By performing such calculations numerically and analytically, we provide priors for the forms of their current distributions and initiate an effort to understand the mechanisms for rare ionic current fluctuations.

\section{\label{sec:main2}Macroscopic fluctuation theory}
For a system whose dynamics are describable by a stochastic conservation equation, macroscopic fluctuation theory provides a way of evaluating the large deviation functions of their associate time-extensive variables. Macroscopic fluctuation theory has been used to evaluate the rate function for current fluctuations, defined as the time-intensive log-likelihood of a time-integrated current.\cite{bertini2015macroscopic} The rate function is evaluated using the so-called contraction principle of large deviation theory.\cite{touchette2009large} This principle is essentially equivalent to a minimum action principle, where the minimization is done over all space-time fields as is done for rate calculations in the instanton limit.\cite{heller2024evaluation,zakine2023minimum} Provided a probability of joint space-time fluctuations of thermodynamic fields and their associate currents, a current rate function is deduced through an optimization procedure where an optimal configuration of the thermodynamic fields is constrained to generate a given current. Traditionally, this is done by formulating a corresponding Euler-Lagrange equation. This theory has been used in lattice models, and recently in models of active matter, uncovering dynamical phase transitions and elucidating non-linear responses.\cite{hurtado2009test,hurtado2011spontaneous,bodineau2005distribution,grandpre2021entropy,fodor2022irreversibility,janas2016dynamical,dolezal2019large,agranov2023macroscopic,jack2020dynamical}

\subsection{\label{sec:main2a}Stochastic Poisson-Nernst-Planck equations}
To apply macroscopic fluctuation theory to an electrolyte solution, we need to first establish the relevant stochastic equations of motion. We consider a solution comprised of a strong aqueous electrolyte, like NaCl, in the dilute limit such that it will be fully dissociated. As a consequence, we assume that interactions between ions are well described by purely Coulomb interactions, neglecting short-ranged interactions important for denser solutions. We further neglect hydrodynamic effects, imagining that the solvent remains quiescent and the friction it imposes on the ions is concentration-independent. Under these assumptions, a Brownian dynamics model can be constructed.\cite{donev2019fluctuating} From those stochastic particle-based equations, we obtain a stochastic variant of the well-known Poisson-Nernst-Planck equations.\cite{dean1996langevin,bonneau2025stationary} The Poisson-Nernst-Planck equations describe the change in time and space of the ionic densities, through a continuity equation that includes both drift and diffusive contributions to ionic currents.\cite{babick2016fundamentals,farhadi2025capacitive}

\begin{figure}[t]
    \centering
    \includegraphics[width=\linewidth]{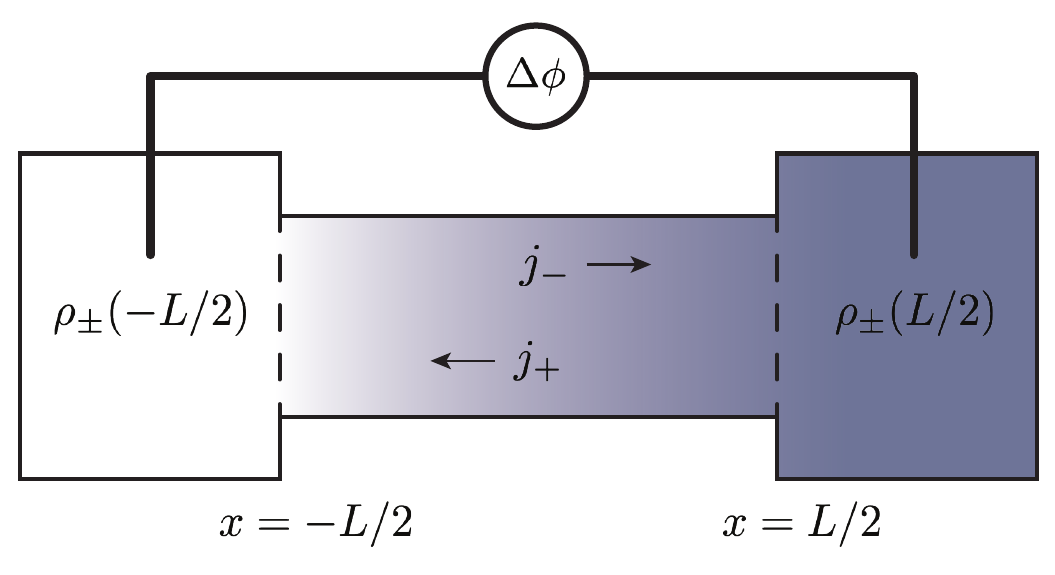}
    \caption{Geometry of the channel we consider, whose long axis lies along the $x$ direction, and whose length is $L$. Boundary conditions fix the densities ${\rho_i(x)}$ at ${x=\pm L/2}$ and the potential drop, ${\Delta \phi}$ and drive currents $j_i$ for ${i=\{+,-\}}$.}
    \label{fig1}
\end{figure}

Throughout, we will consider systems whose time dependence is well described in one dimension like the channel geometry illustrated in Fig.~\ref{fig1}. We will consider two-component electrolytes, whose density fields, ${\rho_i(x,t)}$, depend on the position along the channel, denoted $x$, time $t$, and species index ${i=\{+,-\}}$ for the cation and anion, respectively. The resultant continuity equation is
\begin{equation}
\frac{\partial \rho_i(x,t)}{\partial t} = - \frac{\partial j_i(x,t)}{\partial x} 
\end{equation}
where the current, ${j_i(x,t)}$, for species $i$ is 
\begin{equation}\label{Eq:JPNP}
j_i(x,t)=-D_i \frac{\partial \rho_i(x,t)}{\partial x} - \beta D_i z_i \rho_i \frac{\partial \phi(x,t)}{\partial x} + \xi_i(x,t)
\end{equation}
given by a sum of three contributions. The first term represents mass diffusion driven by density inhomogeneities, with diffusion constant $D_i$. The second term represents the drift of an ion with charge $z_i$ due to an electrostatic potential ${\phi(x,t)}$ with mobility given by an Einstein relation, ${\beta D_i}$, where $\beta$ is the inverse of the temperature times Boltzmann's constant. The electrostatic potential couples the two density fields, ${\rho_+(x,t)}$ and ${\rho_-(x,t)}$, through Poisson's equation,
\begin{equation}\label{eq:pe}
-\varepsilon \frac{\partial^2 \phi(x,t)}{\partial x^2} = z_+ \rho_+(x,t) + z_- \rho_-(x,t)\
\end{equation}
with dielectric constant $\varepsilon$. The final term is a Gaussian noise, whose mean is ${\langle \xi_i(x,t)\rangle=0}$ and is local in time and space, ${\langle \xi_i(x,t)\xi_j(x',t')\rangle=2D_i \rho_i(x) \delta_{ij} \delta(x-x')\delta (t-t')}$.

Because the noise in Eq.~\ref{Eq:JPNP} is Gaussian, the joint probability of the ionic currents and ion densities in space and time, ${P\left(j_+,\rho_+,j_-,\rho_-\right)}$, takes a simple Gaussian form,
\begin{equation}
P\left (j_+,\rho_+,j_-,\rho_- \right ) = \exp \left[ -\Gamma\left (j_+,\rho_+,j_-,\rho_- \right ) \right ]
\end{equation}
where $\Gamma$ is the associated stochastic action,
\begin{equation}\label{eq:MFT}
\Gamma= \sum_{i=\pm} \int_0^\tau dt \int_{-\frac{L}{2}}^{\frac{L}{2}} dx \frac{\left (j_i +D_i \rho'_i + \beta D_i z_i \rho_i \phi' \right )^2}{4 D_i \rho_i}
\end{equation}
which depends parametrically on an observation time $\tau$ and channel domain length $L$. For compactness of notation, we have adopted a prime to denote a spatial derivative, ${\rho_i'(x,t)=\partial \rho_i(x,t)/\partial x}$. We also suppress the arguments of the field variables. Equation~\ref{eq:MFT} is the equivalent of the macroscopic fluctuation theory action, and can be used through marginalization to evaluate the distribution of ionic currents. As a path integral, care must be taken for the proper normalization of ${P \left(j_+,\rho_+,j_-,\rho_-\right)}$ as well as its stochastic calculus interpretation. For the minimum action solutions we consider, neither consideration is relevant.\cite{limmer2024statistical}

\subsection{\label{sec:main2b}Euler-Lagrange equation}
In the limit of long observation times, ${\tau \to \infty}$, for a Markovian equation of motion like Eq.~\ref{Eq:JPNP}, the joint probability of the cation and anion currents, ${P\left(j_+,j_-\right)}$ will take the form
\begin{equation}\label{eq:jrf}
P\left(j_+,j_-\right) = \exp\left[-\tau  I_2\left(j_+,j_-\right)\right]
\end{equation}
which is a consequence of the expected finite correlation times for fluctuations in $j_+$ and $j_-$. The scaling function, $I_2\left(j_+,j_-\right)$, is known as a rate function in the theory of large deviations. It is computable from the contraction principle,\cite{touchette2009large}
\begin{equation}\label{eq:I2}
I_2\left(j_+,j_-\right) =\frac{1}{\tau} \min_{\rho_+,\rho_-}  \Gamma\left (j_+,\rho_+,j_-,\rho_- \right )  
\end{equation}
as a minimization over the conjugate fluctuating density fields. As a minimization, the optimal values of these fields, denoted $\bar{\rho}_i$, are determined by
\begin{equation}\label{eq:min}
\left. \frac{\delta \Gamma}{\delta \rho_+}\right |_{\bar{\rho}_+} = 0 \qquad \left. \frac{\delta \Gamma}{\delta \rho_-}\right |_{\bar{\rho}_-} = 0
\end{equation}
setting the first-order variations of the stochastic action, $\Gamma$, with respect to them, to zero. The contraction principle is analogous to evaluating the marginal distribution of currents using Laplace's method, which is valid because of the large deviation scaling in Eq.~\ref{eq:jrf} that renders all but the most likely fields exponentially suppressed in the long time limit. The rate function is convex,\cite{touchette2009large} with a zero when evaluated at the mean of its arguments, or equivalently to the steady-state solution of non-fluctuating Poisson-Nernst-Planck equations.

The minimization in Eq.~\ref{eq:min} is done over a set of spacetime fields, whose optimums are denoted ${\bar{\rho}_+(x,t)}$ and ${\bar{\rho}_-(x,t)}$. In the long time limit, we assume that the time dependencies vanish due to the development of an associated stationary nonequilibrium steady-state. This assumption can break down if time-periodic states develop instead, which can occur due to the formation of traveling waves under periodic boundary conditions.\cite{hurtado2011spontaneous,bodineau2005distribution}  We consider a system in contact with two reservoirs and will assume that the minimization can be done over a set of time-independent fields. Taking the functional derivatives of the action with respect to variations in the density fields, we obtain equations for the optimal density profiles,   
\begin{equation}\label{eq:ELE}
\left (\frac{j_i}{D_i}\right )^2 = \bar{\rho}_i'^2- 2\bar{\rho}_i \bar{\rho}''_i +\beta z_i \bar{\rho}_i^2 \left( \beta  z_i \phi'^2-2 \phi'' \right)
\end{equation}
for each species $i$, consistent with a given current, $j_i$. The two equations for $\bar{\rho}_+$ and $\bar{\rho}_-$ are coupled through Poisson's equation. This means that Eqs.~\ref{eq:ELE} and Eq.~\ref{eq:pe}, form a set of Euler-Lagrange equations that must be solved simultaneously to determine the rate function ${I_2\left(j_+,j_-\right)}$, or the most likely field configurations for a prescribed current. 

\subsection{\label{sec:main2c}Finite difference solutions}

In general, the Euler-Lagrange equations for ionic currents are coupled, nonlinear second-order differential equations. As a consequence, they will not usually be analytically tractable. In order to study their solutions, we discretize them using a second-order, central, finite difference method.\cite{thomas2013numerical,leveque2007finite} Using a regular grid with uniform spacing  at least one order of magnitude smaller than the Debye screening length, solutions are obtained iteratively using the Newton-Raphson method,
\begin{equation}
    \boldsymbol{u}^{n+1} = \boldsymbol{u}^n - \boldsymbol{J}^{-1} (\boldsymbol{u}^n) \boldsymbol{F}( \boldsymbol{u}^n)
\end{equation}
until ${\| \boldsymbol{u}^{n+1} - \boldsymbol{u}^{n} \| < 10^{-10}}$ where $\boldsymbol{u}^n$ is the vector comprising the discretized approximation to $\bar{\rho}_i$ and $\phi$, and $\boldsymbol{F}$ is the finite difference approximation to Eqs.~\ref{eq:ELE} and~\ref{eq:pe}. The tensor ${\boldsymbol{J} = \partial \boldsymbol{F} / \partial \boldsymbol{u}}$ is Jacobian matrix. We use an initial guess for $\boldsymbol{u}^0$ that is the linear profile for ${j_+ = \langle j_+ \rangle}$ and ${j_- = \langle j_- \rangle}$ and we march spirally in the ${(j_+, j_-)}$ space of interest and use each optimal profile as the initial guess for the next one set of parameters. Having the optimal profiles, we can compute the action from Eq.~\ref{eq:MFT} and then the rate function from Eq.~\ref{eq:I2} using a trapezoidal rule for the integral.

\section{\label{sec:main3}Ionic current fluctuations}
While the expressions for the Euler-Lagrange equations and rate functions are general, here we will focus on the simplest case of a fully symmetric electrolyte, where ${z_+=-z_-=z}$ and ${D_+=D_-=D}$. As a consequence, it is useful to define a charge density ${q=z\left(\rho_+-\rho_-\right)}$, and associated ionic current ${j_q=z\left(j_+-j_-\right)}$, as well as a mass density field, ${\rho=\rho_++\rho_-}$, and associated mass current ${j_\rho=j_++j_-}$.  As three coupled second-order differential equations, our Euler-Lagrange equations require six boundary conditions. For the open channel geometry we consider here, we will restrict our attention to currents driven by a potential drop, and apply boundary conditions ${\rho\left(\pm L/2\right)=\hat{\rho}}$ and ${q\left(\pm L/2\right)=0}$, while ${\phi\left(\pm L/2\right)=\pm \Delta \phi/2}$.

Written in terms of the charge density, mass density, and their respective currents, Eqs.~\ref{eq:ELE} become
\begin{align}
\begin{split}
    \frac{j_q j_\rho}{D^2} =&  4\beta^2 z^2 \bar{q}\bar{\rho} \phi'^2 - \beta \left (\bar{q}^2+ z^2 \bar{\rho}^2\right ) \phi'' \\
    &+ \bar{q}'\bar{\rho}' - \bar{q}''\bar{\rho} - \bar{q}\bar{\rho}'' 
\end{split}
\end{align}
and 
\begin{align}
\begin{split}
    \frac{j_q^2 +z^2 j_\rho^2}{D^2 } &= \bar{q}'^2 - 2\bar{q}\bar{q}''+ z^2\left(\bar{\rho}'^2 - 2\bar{\rho}\bar{\rho}''\right) \\
    &\quad\, + \beta^2 z^2\left( \bar{q}^2 + z^2 \bar{\rho}^2 \right) \phi'^2 - 4 \beta z^2 \bar{q}\bar{\rho} \phi''
\end{split}
\end{align}
which we arrive at by eliminating $\bar{\rho}_\pm$ and $j_\pm$ and taking sums and differences of the two resulting equations.
Correspondingly, Poisson's equation is written solely in terms of the charge density, ${\phi''(x) = -q(x)/\varepsilon}$. 

From the joint rate function in Eq.~\ref{eq:I2}, the rate function for an ionic current, ${I(j_q)=-\tau^{-1}\ln P(j_q)}$  can be evaluated from an analogous contraction principle,
\begin{equation}\label{eq:minrq}
I(j_q) = \min_{j_\rho}  I_2\left (j_q,j_\rho\right ) 
\end{equation}
by minimizing over $j_\rho$. For the symmetric electrolyte we consider and for currents generated only by a potential drop, the mean mass current is zero. Further, under these conditions, ${I_2\left(j_q,j_\rho\right)=I_2\left(j_q,-j_\rho\right)}$. As a consequence of this symmetry and the convexity of the rate function, we find that the minimization over $j_\rho$ enforces that ${\bar{q}=0}$. This significantly simplifies the Euler-Lagrange equations. We find that the equations and their resultant rate functions are analytically tractable to solve in limiting cases of small and large applied potential. In the following, we compare those limiting expressions to those deduced from numerical solutions to the full equations. 

\subsection{\label{sec:main3a}Small applied potential}

For the symmetric electrolyte with a potential drop, the minimization over mass currents yields ${\bar{q}=0}$ and ${j_\rho=0}$. From Poisson's equation, this implies that the potential profile is linear, ${\phi(x)=x\Delta \phi /L}$. In the limit that ${\Delta\phi=0}$, the Euler-Lagrange equation that determines the likelihood of an ionic current reduces to
\begin{equation}
\left(\frac{j_q}{D z}\right)^2 = \bar{\rho}'^2- 2 \bar{\rho} \bar{\rho}''
\end{equation}
which is a quadratic, second-order differential equation. This equation can be solved, yielding a parabolic profile,
\begin{equation}
\frac{\bar{\rho}(x)}{\hat{\rho}}= 1 + \frac{1}{2} \left (\frac{j_q L}{2Dz\hat{\rho}} \right )^2 \frac{1-4(x/L)^2}{1+ \sqrt{1+(j_q L/2Dz\hat{\rho})^2}} 
\label{eq:pxnophi}
\end{equation}
whose amplitude depends on the value of the current, naturally expressed in units of ${2Dz\hat{\rho}/L}$, and is symmetric about ${x=0}$. The reflection symmetry is a consequence of the inversion symmetry of the system.

\begin{figure}[t]
    \centering
    \includegraphics[width=\linewidth]{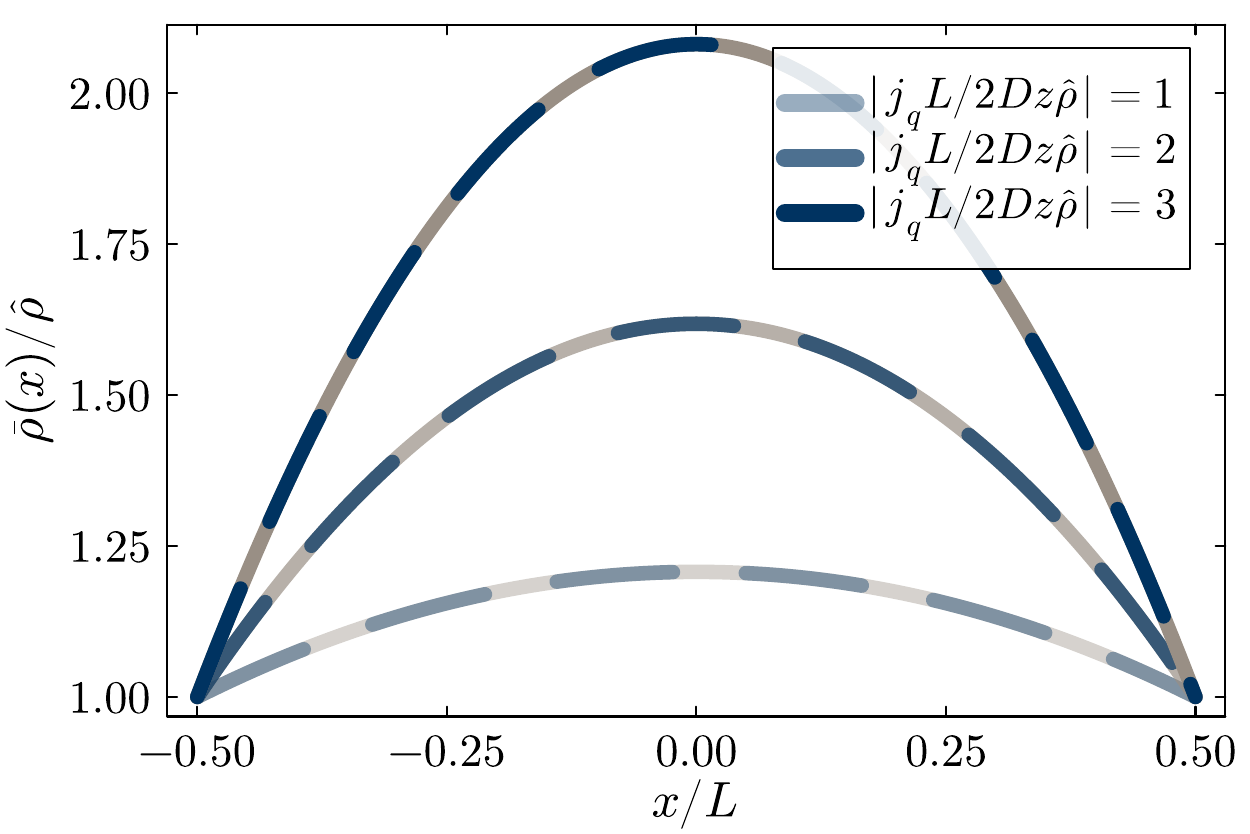}
    \caption{
    Optimal mass density profile for ${\Delta\phi = 0}$. Solid lines are the results of finite difference simulations. Dashed lines are the analytical solution in Eq.~\ref{eq:pxnophi}.}
    \label{fig:rholo}
\end{figure}

Figure \ref{fig:rholo} compares the analytical solution in Eq.~\ref{eq:pxnophi} to that obtained by the finite difference solution to Eqs.~\ref{eq:ELE}. The agreement is exact, confirming our formal inferences concerning the minimization over $j_\rho$. For different values of the conditioned current, the profiles obtain different heights, increasing the curvature around ${x=0}$. The maximum height of the density fluctuation scales quadratically with $j_q$ for small values, ${|j_q|< 2D z \hat{\rho}/L}$, and eventually linearly with $|j_q|$. This implies that the most likely way to generate an ionic current fluctuation in this channel is to create a density inhomogeneity that is delocalized over the length of the channel. Similar optimal profiles have been found in cases of passive diffusion of dilute solutes.\cite{bodineau2004current}

From the optimal density profile, the ionic current rate function can be evaluated by integrating the remaining action. The rate function is expressible as
\begin{equation}
I(j_q) = \frac{2D \hat{\rho}}{L} \, \mathcal{I} \left (\frac{j_q L}{2Dz \hat{\rho}} \right)
\end{equation}
with an amplitude ${2D \hat{\rho}/L}$ and dimensionless scaling function
\begin{equation}
\mathcal{I}(j) = 1-\sqrt{1 + j^2} + 2j \tanh^{-1} \left(\frac{j}{1+\sqrt{1+j^2}}\right)
\end{equation}
where ${\tanh^{-1}(y)=\ln(\sqrt{1+y})-\ln(\sqrt{1-y})}$ is the inverse of the hyperbolic tangent function. Notably, ${\mathcal{I}(j)}$ is not quadratic, so the distribution of ionic currents is not Gaussian. For small values of its argument, ${\mathcal{I}(j) \approx j^2/2 +\mathcal{O}(j^4)}$ is Gaussian, as expected from the central limit theorem and consistent with time reversal symmetry, ${I(j_q)=I(-j_q)}$. The variance of those Gaussian fluctuations is ${\beta \tau \langle \delta j_q^2 \rangle = 2D \kappa^2\varepsilon/L}$, where ${\kappa=\sqrt{\beta z^2 \hat{\rho}/\varepsilon}}$ is the inverse Debye screening length. The variance is equal to twice the Nernst-Einstein conductivity times ${k_\mathrm{B} T}$.\cite{onsager1953fluctuations} The ionic current distribution has exponential tails, as asymptotically for large magnitudes of $j_q$, ${\mathcal{I}(j)\sim |j| (\ln|2j|-1)}$. The exponential tail results from the rare currents being driven convectively by the development of the density inhomogeneity, rather than by a sum of uncorrelated thermal fluctuations, which is more likely for small fluctuations.

\begin{figure}[t]
    \centering
    \includegraphics[width=\linewidth]{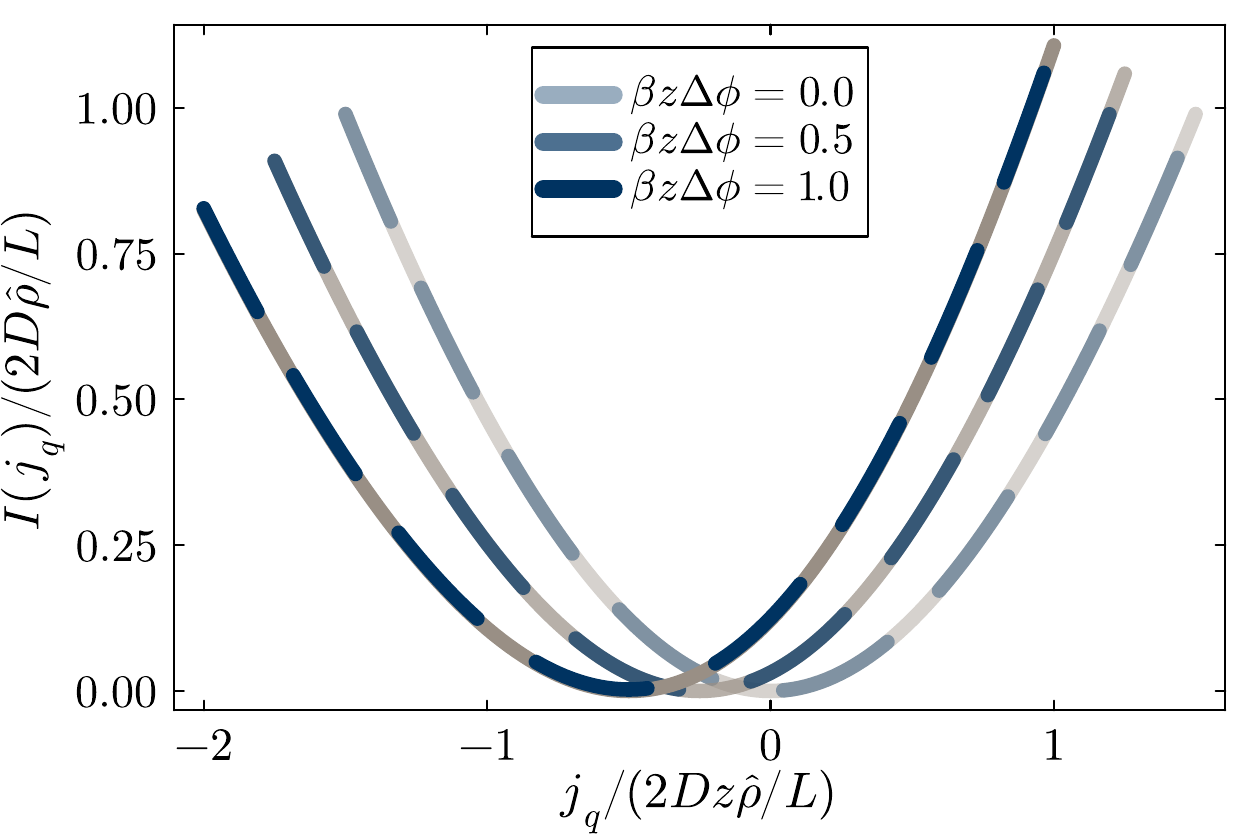}
    \caption{Rate function for ionic currents in the limit of low applied potential. Solid lines are the results of finite difference simulations. Dashed lines are the perturbative analytical solution in Eq.~\ref{eq:IqPT}.}
    \label{fig:Iqlp}
\end{figure}

For small values of the applied potential, ${|\beta z \Delta \phi| \ll 1}$, we can evaluate the rate function perturbatively using the zero potential density profiles as a reference. Using the density profile in Eq.~\ref{eq:pxnophi} for the integrals in the expression for the stochastic action,  the rate function to second order in $\Delta\phi$ is
\begin{align}\label{eq:IqPT}
\begin{split}
I(j_q) &\approx \frac{2D \hat{\rho}}{L} \, \mathcal{I} \left (\frac{j_q L}{2Dz \hat{\rho}} \right )+ \frac{\beta j_q \Delta \phi}{2}  \\
&\quad\, + \frac{ D\beta^2 z^2 \Delta \phi^2 \hat{\rho}}{12 L} \left[ 2+\sqrt{1+\left(\frac{j_q L}{2 D z \hat{\rho}}\right )^2} \right]
\end{split}
\end{align}
which adds to the case of ${\Delta\phi=0}$ a term bilinear in $j_q$ and $\Delta\phi$, and a nonlinear term proportional to the square root of the squared current. Rate functions for finite, small applied potential are shown in Fig.~\ref{fig:Iqlp}. The addition of the applied potential shifts the minimum of the rate function. 
The mean current can be computed by finding the location of this minimum of ${I(j_q)}$. To first order in the applied potential, ${\langle j_q \rangle \approx - D \kappa^2 \varepsilon \Delta \phi/L}$ in agreement with the direct solution of the Poisson-Nernst-Planck equations. This is also consistent with the fluctuation-dissipation theorem, ${\langle j_q \rangle = -\beta \tau \langle \delta j_q^2 \rangle \Delta \phi/2}$, provided the zero applied potential variance of the ionic current. For ${|\beta z \Delta \phi| <1}$, the perturbative approximation to ${I(j_q)}$ in Eq.~\ref{eq:IqPT} is accurate, but beyond this regime, its accuracy degrades. 

\subsection{\label{sec:main3b}Large applied potential}

In the limit of large applied potentials, it is useful to define dimensionless variables, ${\alpha = \beta z \Delta \phi/2}$, ${\hat{j}=j_q/\langle j_q \rangle}$ where ${\langle j_q \rangle=-D \kappa^2 \varepsilon \Delta \phi/L}$. Written in terms of these variables, the Euler-Lagrange equation becomes,
\begin{equation}\label{eq:ELjq}
\hat{j}^2 = \left(\frac{\bar{\rho}}{\hat{\rho}}\right)^2 + \left( \frac{L}{2\alpha}\right)^2 \left( \frac{\bar{\rho}'^2 - 2\bar{\rho} \bar{\rho}''}{\hat{\rho}^2} \right)
\end{equation}
which restores the contribution quadratic in the density that was neglected in the low-potential case. The optimal density profile in this case has general solutions of the form,
\begin{equation}
\frac{\bar{\rho}(x)}{\hat{\rho}} = \frac{e^{-2 \alpha x/L}}{16\alpha^2 c_2} \left [\left ( c_2 e^{2 \alpha x/L} - c_1\right )^2 - 64 \alpha^4 \hat{j}^2 \right]
\end{equation}
where $c_1$ and $c_2$ are constants of integration fixed by the boundary conditions.

\begin{figure}[t]
    \centering
    \includegraphics[width=\linewidth]{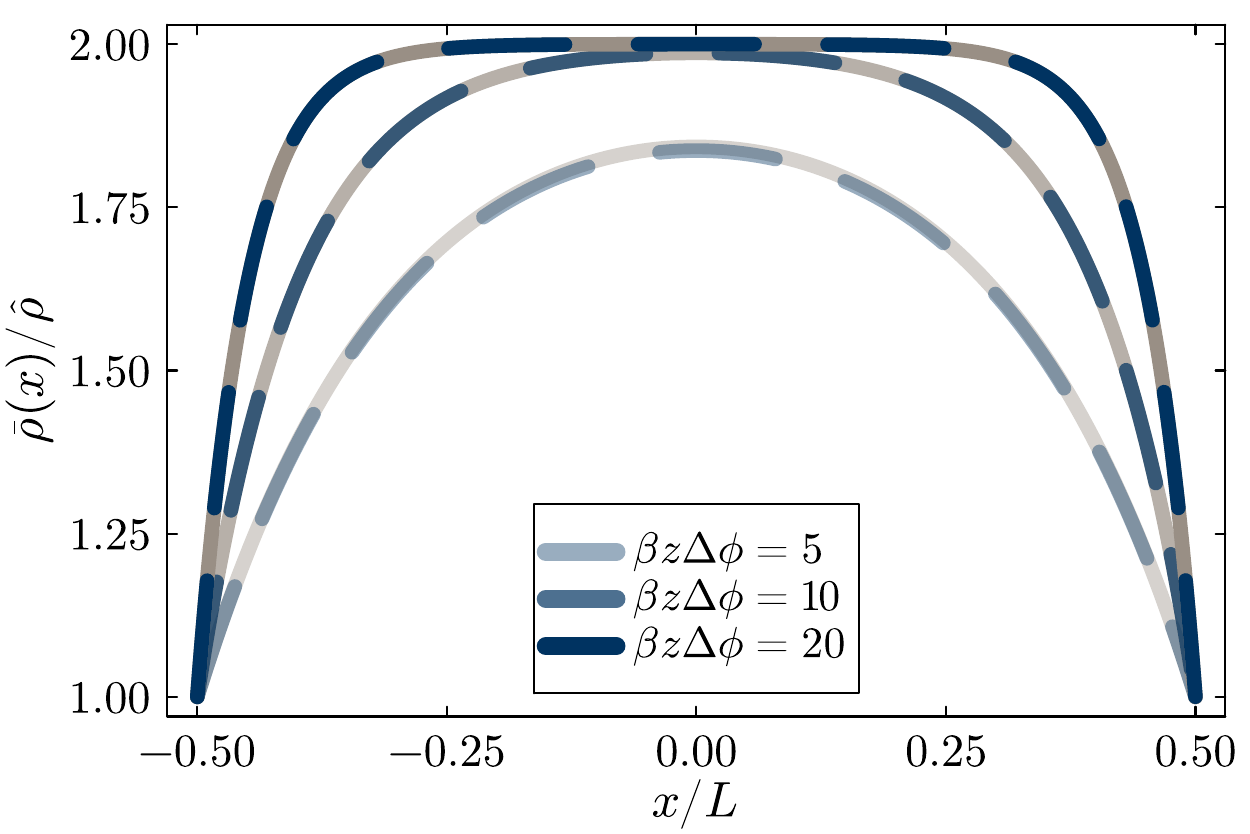}
    \caption{
    Optimal mass density profile for large $\beta z \Delta\phi$ at fixed ${|j_q / \langle j_q \rangle| = 2}$. Solid lines are the results of finite difference simulations. Dashed lines are the approximate analytical solution in Eq.~\ref{eq:pxhp}}
    \label{fig:rhohp}
\end{figure}

Solving the boundary value problem above is cumbersome. In the limit that ${\alpha^2 \gg 1}$, we find a simple limiting form
\begin{equation}\label{eq:pxhp}
\frac{\bar{\rho}(x)}{\hat{\rho}} \approx |\hat{j}|+\left (1-|\hat{j}| \right ) \frac{\cosh(2\alpha x/L)}{\cosh(\alpha)}
\end{equation}
whose amplitude increases linearly for ${|j_q| > |\langle j_q \rangle|}$, and decreases for ${|j_q| < |\langle j_q \rangle|}$. Unlike the profiles for zero applied potential, these optimal density profiles have a characteristic length, ${L/\beta z |\Delta \phi|}$, interpretable as a field penetration depth.\cite{palaia2023poisson} For distances into the channel greater than this penetration depth, the density profile is flat. Thus, in the case of large applied potentials, the mechanism of generating rare current fluctuations is the generation of boundary layers that alter the density of ions uniformly through the channel, forcing the ions to convect and generate a current. The approximate optimal density profiles are compared to the numerically exact solutions in Fig.~\ref{fig:rhohp}. For ${\beta z |\Delta\phi| > 1}$ we find that the density profiles are accurate, while below this value, they are better described by the zero potential, parabolic profile.

Using the approximate optimal density profile in Eq.~\ref{eq:pxhp}, the ionic current rate function can be obtained by integrating over the resultant action. We find for large applied potentials,
\begin{align}\label{eq:ratehp}
\begin{split}
I(j_q) =& \frac{2D \hat{\rho} \alpha}{L} \left [-\alpha \hat{j} - (|\hat{j}|-1) \tanh(\alpha) \right. \\
&+ \left . 2 \hat{j}^2 \frac{\cosh(\alpha) \tanh^{-1}(\Theta)}{\sqrt{\hat{j}^2\cosh^2(\alpha)-(|\hat{j}|-1)^2}} \right]
\end{split}
\end{align}
where
\begin{equation}
\Theta=\frac{|\hat{j}|\sinh(\alpha)-\tanh(\alpha/2)}{\sqrt{\hat{j}^2\cosh^2(\alpha)-(|\hat{j}|-1)^2}}
\end{equation}
which contain non-analytic contributions from the absolute value of the current around ${\hat{j}=0}$.  As before, the rate function has a natural scale, given by ${2D\hat{\rho}/L}$, times a dimensionless scaling function that now is expressed in terms of the current relative to its mean and the dimensionless applied potential. 

\begin{figure}[b]
    \centering
    \includegraphics[width=\linewidth]{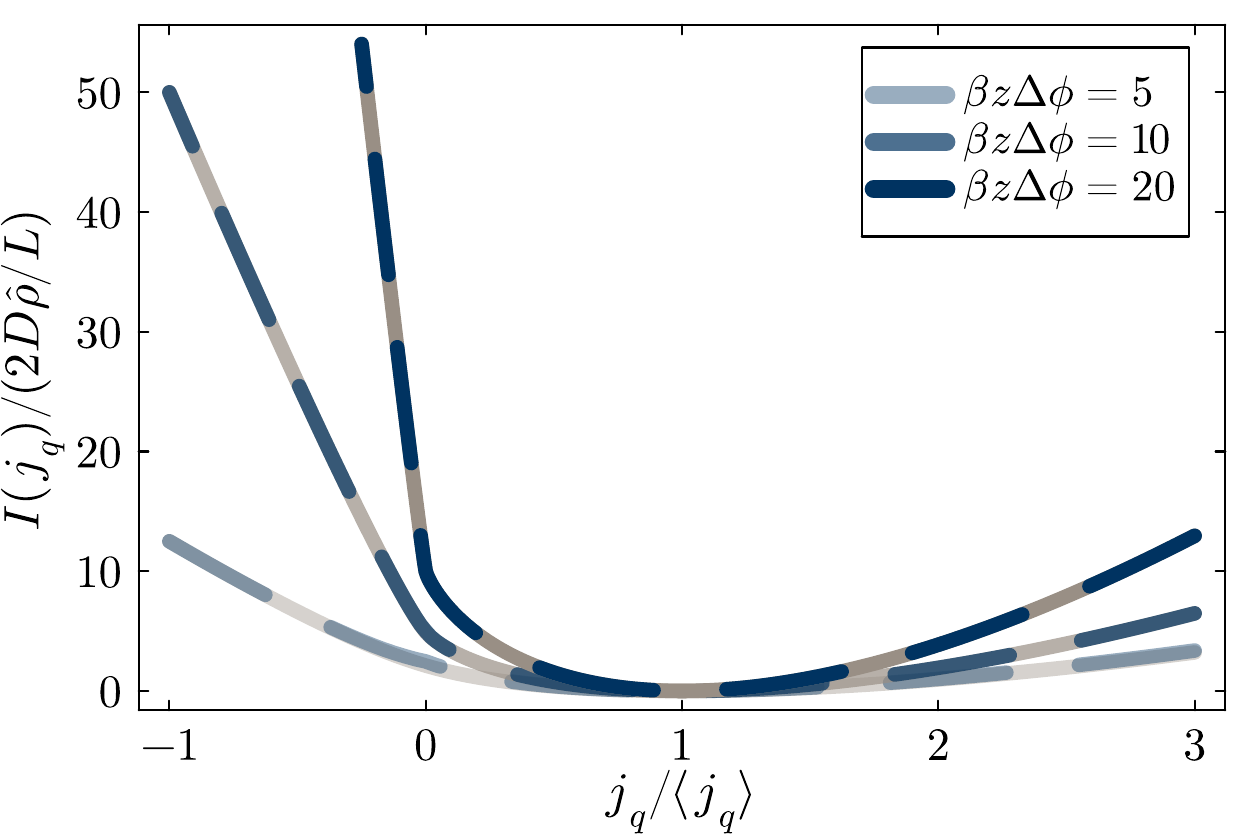}
    \caption{Rate function for ionic currents for large applied potential. Solid lines are the results of finite difference simulations. Dashed lines are the approximate analytical solution in Eq.~\ref{eq:ratehp}. }
    \label{fig:ratehp}
\end{figure}

Exact numerical results for ${I(j_q)}$ are compared to Eq.~\ref{eq:ratehp} in Fig.~\ref{fig:ratehp}. For the values of the applied potential considered, the correspondence is very good, indistinguishable to graphical accuracy. The rate function is manifestly non-Gaussian, and highly asymmetric. Fluctuations to larger than average current occur with relatively high probability, while fluctuations in the current to directions opposite of its mean are sharply suppressed. These currents go in opposition to those expected from the average increase of the entropy, and the resultant asymmetry is a consequence of the breaking time reversal symmetry by driving a finite mean current through the system. We explore this point further below.\cite{limmer2024statistical}

\section{\label{sec:main4}Thermodynamics of ionic currents}
The development of a persistent current through a system requires the constant injection of energy. As a consequence, thermodynamic constraints on typical currents can be formulated as direct implications of the second law. For example, the second law requires that entropy production of a system in steady-state is non-negative, so that the product of a current and its thermodynamic driving force is nonnegative.\cite{de2013non} Stochastic thermodynamic generalizations on the second law have been analogously formulated to constrain fluctuations of currents away from their typical, mean value. These include fluctuation theorems that impose specific symmetry relations between currents and the reversed values at the level of the rate function.\cite{gallavotti1995dynamical,lebowitz1999gallavotti,jarzynski2011equalities,sevick2008fluctuation} Additionally, thermodynamic uncertainty relations have recently been derived that constrain ratios of moments of the current distribution to the total entropy production in the system.\cite{horowitz2020thermodynamic,seifert2019stochastic} With access to the rate function for ionic currents, we explore both of these thermodynamic constraints below.

\subsection{\label{sec:main4a}Gallavotti-Cohen symmetry}
Gallavotti-Cohen symmetry is an example of a fluctuation theorem that relates the probabilities of a given current fluctuation and its inverse, to the rate of entropy production generated by that current, $\dot{\sigma}$.\cite{gallavotti1995dynamical} It relates symmetries that hold generally for dynamical systems whose equations of motion are thermodynamically consistent, and implies the subtle consequences of time-reversal symmetry of microscopic dynamics to the macroscopic, irreversible emergent behavior. For the joint rate function of cation and anion currents, we find that it satisfies Gallavotti-Cohen symmetry for arbitrary density profiles,
\begin{equation}
I_2\left(j_+,j_-\right) - I_2\left(-j_+,-j_-\right) = \beta j_+ \Delta \mu_++ \beta j_- \Delta \mu_-
\end{equation}
where ${\beta\mu_\pm (x)= \ln [\rho_\pm(x) e^{\beta z_\pm \phi(x)}]}$ is the model for the electrochemical potential of the ions within the approximations of the Poisson-Nernst-Planck equations and ${\Delta \mu_\pm = \mu_\pm(L/2)-\mu_\pm(-L/2)}$. This model includes contributions from the ideal chemical potential as well as the electrostatic self-energy of the charge under the spatially varying potential ${\phi(x)}$. Written in terms of the individual ion mass currents, the entropy production is bilinear in the mass current and the respective electrochemical potential drop across the channel. Written in terms of the mass and ionic currents for a symmetric electrolyte, this implies
\begin{align}
\begin{split}
I(j_q,j_\rho)&-I(-j_q,-j_\rho) = \beta j_q \Delta \phi \\
&+\frac{j_q}{2z} \Delta \ln \left (\rho_+ /\rho_- \right) + \frac{j_\rho}{2} \Delta \ln \left( \rho_+ \rho_- \right)
\end{split}
\end{align}
namely, an analogous symmetry where the entropy production is generically given by a sum of three terms. The first is the ionic current times the potential drop. The second and third reflect the ability for mass and charge imbalances at the boundaries to drive currents. 

\begin{figure}
    \centering
    \includegraphics[width=\linewidth]{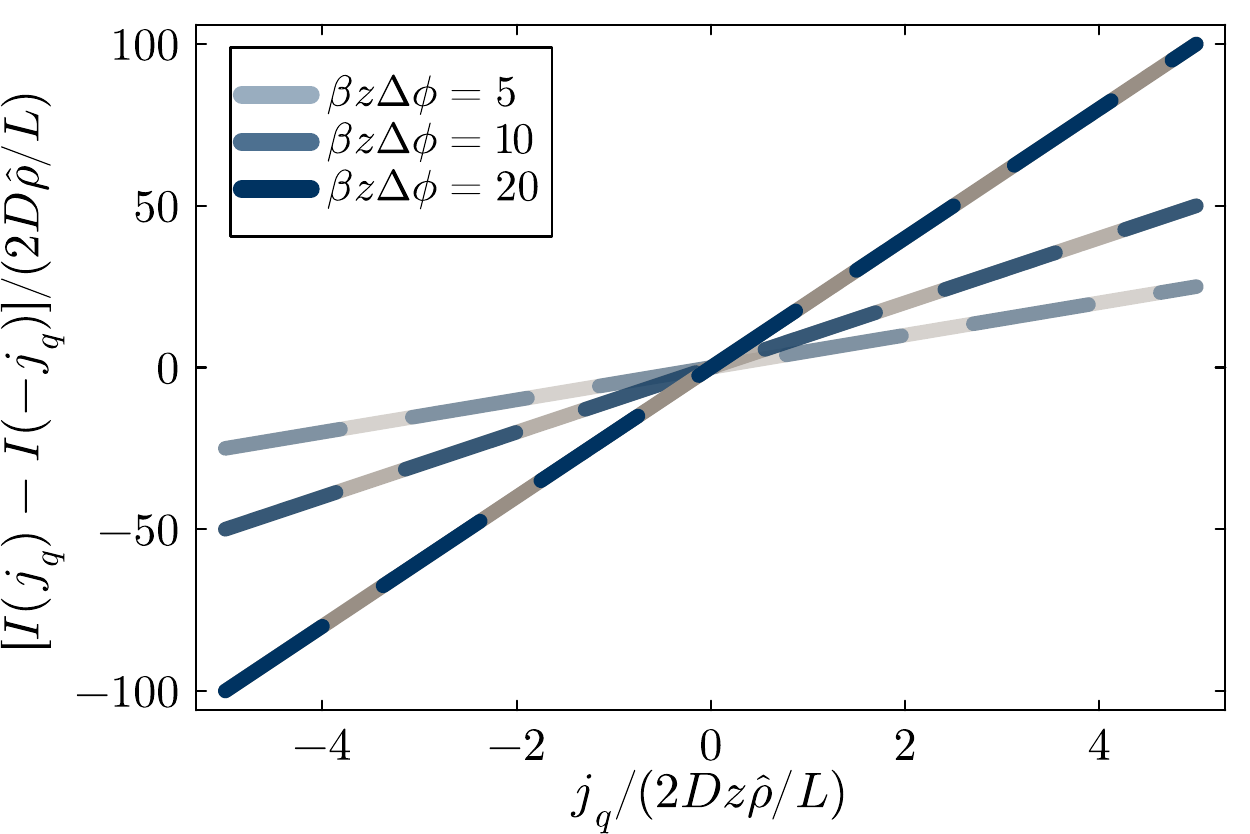}
    \raggedleft
    \includegraphics[width=0.95\linewidth]{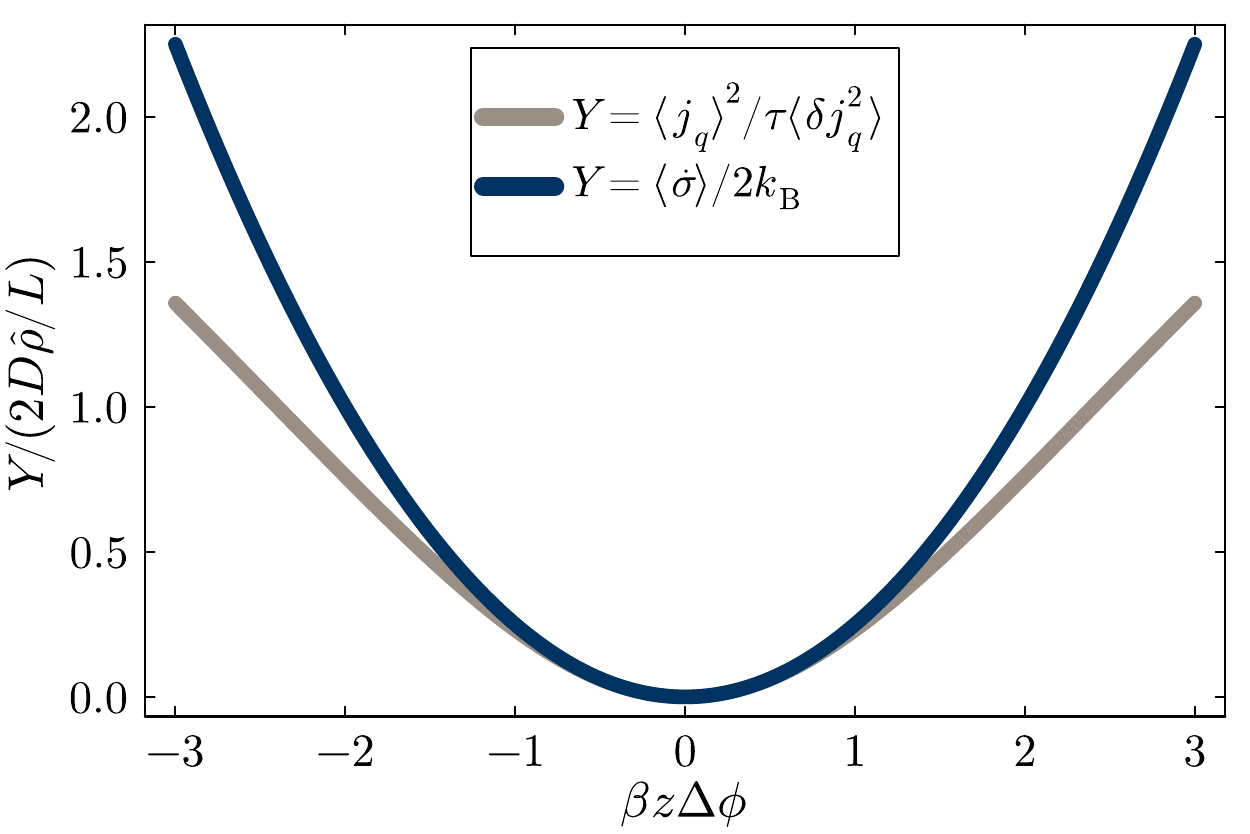}
    \caption{(top) Illustration of Gallavotti-Cohen symmetry for the rate functions in Fig.~\ref{fig:ratehp}. (bottom) Thermodynamic uncertainty relation for the ionic currents driven by a potential drop.}
    \label{fig:GC}
\end{figure}

For the rate function for solely ionic currents, under the boundary conditions we consider in Sec.~\ref{sec:main3}, the Gallavotti-Cohen symmetry reduces to 
\begin{equation}
I(j_q)-I(-j_q)=\beta j_q \Delta \phi
\end{equation}
which implies an average rate of entropy production,
\begin{equation}
\dot{\sigma}/k_\mathrm{B}=-\beta j_q  \Delta \phi
\end{equation}
which identifies ${-\Delta\phi}$ as the relevant thermodynamic force for driving ionic currents. Both the perturbative solution for the rate function for small applied potentials in Eq.~\ref{eq:IqPT} and the approximate solution for large applied potentials in Eq.~\ref{eq:ratehp} satisfy this symmetry. Gallavotti-Cohen symmetry is thus the origin of the sharp linear feature that suppresses current fluctuations in the direction negative to the mean current, as illustrated in Fig.~\ref{fig:GC}.

\subsection{\label{sec:main4b}Fluctuation-response relations}
Gallavotti-Cohen symmetry implies a relationship between moments of the current distribution known as the thermodynamic uncertainty relation.\cite{hasegawa2019fluctuation} This principle states that the certainty with which the current can be measured, defined as the ratio of the mean current squared divided by its variance multiplied by the observation time, ${\langle j_q \rangle^2/\tau\langle \delta j_q^2\rangle}$, is bounded from above by a function of the rate of entropy produced in the system. In certain limits, this function is simply linear in the rate of entropy production.\cite{barato2015thermodynamic} We find that for ionic currents described by the fluctuating Poisson-Nernst-Planck equations, this relation is satisfied. 

From either the direct solution of the Poisson-Nernst-Planck equations or from our rate functions, we can deduce the mean current driven by a potential drop,
\begin{equation}
\langle j_q \rangle = -D  \kappa^2 \varepsilon \Delta \phi/L
\end{equation}
which is a product of the Nernst-Einstein conductivity and minus the gradient of the potential. From Gallavotti-Cohen symmetry, the mean rate of entropy production is then
\begin{equation}
\langle \dot{\sigma}\rangle /k_\mathrm{B} = \beta D  \kappa^2 \varepsilon \Delta \phi^2/L 
\end{equation}
or the current times the negative gradient of the potential across the channel, where $k_\mathrm{B}$ is Boltzmann's constant. We can evaluate the potential drop dependent variance in the current by linearizing Eq.~\ref{eq:ELjq} about a constant density profile. The subsequent profile and rate function can be evaluated and the variance,
\begin{equation}
\tau \langle \delta j_q^2 \rangle = \frac{D  \kappa^2 \varepsilon}{\beta L}  \frac{ \beta z \Delta \phi }{ \tanh(\beta z \Delta \phi/2)}
\end{equation}
determined by expanding the rate function around its average to second order. While the mean current is a simple linear function of the applied potential, the variance of the current depends strongly on the applied potential.

Assembling these results, we find that the Poisson-Nernst-Planck equations satisfy a strong thermodynamic uncertainty relation,
\begin{equation}
\frac{\langle \dot{\sigma}\rangle }{2 k_\mathrm{B}} \geq \frac{\langle j_q\rangle^2}{\tau \langle \delta j_q^2 \rangle } 
\end{equation}
where the certainty of the current is bounded from above by ${\langle \dot{\sigma}\rangle /2 k_\mathrm{B}}$. This relation is illustrated in Fig~\ref{fig:GC}, where we find the bound is saturated for ${|\beta z \Delta \phi| < 1}$, which constitutes the domain of linear response where the fluctuation-dissipation relationship is valid. Beyond that regime, we find by rearranging the thermodynamic uncertainty relationship,
\begin{equation}
\frac{\beta \tau}{2} \langle \delta j_q^2 \rangle |\Delta\phi| \geq |\langle j_q \rangle|
\end{equation}
that the linear response estimate of the mean current (the product of the variance and applied potential) becomes much larger than the actual observed mean current. While the certainty of the current increases with applied potential beyond ${|\beta z\Delta\phi| \approx 1}$, the energy consumption required to maintain that steady-state is much larger, so taking the thermodynamic uncertainty principle as a measure of efficiency, large applied potentials are inefficient at producing low-noise currents. 

\section{\label{sec:main5}Conclusion}
Using macroscopic fluctuation theory, we have established the distribution of ionic currents for a strong, dilute electrolyte in a one-dimensional channel under an applied potential. This is possible by describing the dynamics of the ion density field with a stochastic version of the Poisson-Nernst-Planck equation and formulating an Euler-Lagrange equation for the most likely density profile conditioned on a given current. Focusing on a symmetric electrolyte, we have found that generically the fluctuations of the ionic current are non-Gaussian. The mechanisms for generating rare fluctuations depend on the size of the applied potential. While we have considered simple model electrolytes in simple geometries in order to derive analytic results, the finite difference solutions to the Euler-Lagrange equations can be easily extended to non-symmetric electrolytes, complicated geometries, and boundary conditions where mass currents are generated in addition to ionic currents. By parameterizing concentration-dependent mobilities, this same formalism can be extended to more concentrated electrolytes, opening the potential for time-dependent optimal profiles, and dynamical phase transitions.

\section*{ACKNOWLEDGMENTS}
This work was supported by the Condensed Phase and Interfacial Molecular Science Program (CPIMS) of the U.S. Department of Energy under contract no. DE-AC02-05CH11231. The authors are grateful to Kranthi K. Mandadapu and Karthik Shekhar for their support throughout this work.

\appendix
\renewcommand{\thefigure}{A\arabic{figure}}
\setcounter{figure}{0}

\section*{References}
\bibliographystyle{aipnum4-2}
\bibliography{main}

\end{document}